\documentclass[5p,authoryear]{elsarticle}

\usepackage{amsmath}
\usepackage{amssymb}

\usepackage[varg]{txfonts}
\usepackage{color,soul}

\newcommand{\OrderOf}[1]{\ensuremath{{\mathcal O}\left(#1\right)}}
\newcommand{\degree}{\ensuremath{^\circ}}
\newcommand{\degC}{\ensuremath{\degree\mathrm{C}}}
\newcommand{\Figref}[1]{Figure~\ref{#1}}
\newcommand{\Figsref}[1]{Figures~\ref{#1}}
\newcommand{\Tabref}[1]{Table~\ref{#1}}
\newcommand{\Eqref}[1]{Eq.~\eqref{#1}}

\newcommand{\Rey}{\ensuremath{\mathrm{Re}}}
\newcommand{\Pe}{\ensuremath{\mathrm{Pe}}}
\newcommand{\Kb}{\ensuremath{K_{\mathrm{b}}}}

\journal{Chemistry and Physics of Lipids}

\begin{document}

\begin{frontmatter}


  \title{Influence of Micro-mixing on the Size of Liposomes
    Self-Assembled from Miscible Liquid Phases}

\author{Sopan M. Phapal}
\ead{sopan.mp@gmail.com}
\author{P. Sunthar\corref{ps}
}
\ead{P.Sunthar@iitb.ac.in}
\cortext[ps]{Corresponding author}

\address{Department of Chemical Engineering, Indian Institute of
  Technology Bombay (IITB), Powai, Mumbai--400076. India.}

\begin{abstract}
  Ethanol injection and variations of it are a class of methods where
  two miscible phases---one of which contains dissolved
  lipids---are mixed together leading to the self-assembly of lipid
  molecules to form liposomes. This method has been suggested, among
  other applications, for in-situ synthesis of liposomes as drug
  delivery capsules.  However, the mechanism that leads to a specific
  size selection of the liposomes in solution based self-assembly in
  general, and in flow-focussing microfluidic devices in particular,
  has so far not been established.  Here we report two aspects of this
  problem. A simple and easily fabricated device for synthesis of
  monodisperse unilamellar liposomes in a co-axial flow-focussing
  microfluidic geometry is presented.  We also show that the size of
  liposomes is dependent on the extent of micro-convective mixing of
  the two miscible phases.  Here, a viscosity stratification induced
  hydrodynamic instability leads to a gentle micro-mixing which
  results in larger liposome size than when the streams are mixed
  turbulently.  The results are in sharp contrast to a purely
  diffusive mixing in macroscopic laminar flow that was believed to
  occur under these conditions. Further precise quantification of the
  mixing characteristics should provide the insights to develop a
  general theory for size selection for the class of ethanol injection
  methods.  This will also lay grounds for obtaining empirical
  evidence that will enable better control of liposome sizes and for
  designing drug encapsulation and delivery devices.

\end{abstract}

\end{frontmatter}

\section{Introduction}

Liposomes offer a great potential in drug delivery systems
\citep{Lasic1992279} in terms of versatility in size and charge,
ability to encapsulate hydrophilic (inside the aqueous core) and
lipophilic drugs (within the lipid bilayer), relative non-toxicity
compared to other carrier systems, ability to escape macrophages in
the lungs (stealth liposomes) and ease of preparing covalently coupled
antibodies to ensure their cell specificity i.e., targeting
\citep{Torchilin2005145, Lasic1998307}. Liposome size and size
distribution are critical parameters to control, since these influence
drug dosage and the clearance rate of the drug from the body. Liposome
sizes in the range 100~nm to 200~nm is most suitable for passive
targeting of cancerous tissue by enhanced permission and retention
mechanism (EPR) \citep{Maeda2000271,Lembo201053}.

One of the ways liposome formation occurs is when of lipids molecules
self-assemble from a solution phase due to changed solvent conditions,
such as increase in hydrophobic costs due to the entry of water. This
phenomenon may be classified under a broad principle of coacervation
\citep{Ishii1995483} and ethanol injection is one such coacervation
methods in which lipids dissolved in ethanol are brought into contact
with water \citep{Batzri19731015, Kremer19773932, Wagner2002259} .
Ethanol and water being miscible in all proportions, leads to a quick
mixing of the phases resulting in induction of the lipid self-assembly
into bilayers. The initial structures are disc-like membranes, which
grow and close upon themselves or fuse with other discs to form
vesicular compartments. The size of these lipid vesicles ranges from
about 20~nm up to about a few micrometers and they may be composed of one or
more concentric membranes (unilamellar and multilamellar vesicles),
each with a thickness of about 4~nm~\citep{Lasic1993}.

There is a fair understanding of the phenomenon of liposome formation
in spontaneous self-assembly---formation of discs of the bilayer
membrane \citep{Lasic1992279} which close upon to form vesicles after
a critical size when the edge energy balances the bending energy
\citep{Helfrich1973693} and in mixed amphiphile systems which
introduce a spontaneous curvature
\citep{Safranetal91,Jungetal2001,Leng20031624}.
However, there has so far been no definite prediction of the size of
liposomes formed in the ethanol injection methods, mainly owing to
method of mixing the two phases and a wide distribution of sizes. The
size prediction and control, as outlined above, is crucial from a
drug-delivery point of view.

In the general class of ethanol injection methods suggested in the
literature is the microfluidic method, offering a fine control on
manipulation of fluid and interfaces, and in small volumes of fluids
in the picoliter range. Recently, \citet{Jahn20042674, Jahn20076289}
introduced a microfluidic two stream hydrodynamic-focussing method
\citep{Ottino2004923} which was shown to generate liposomes of varying
sizes and distributions by controlling the flow rates.  In this
device, phospholipids dissolved in a water miscible solvent (such as
ethanol) is sent through central microchannel and water through a side
microchannel in a planar geometry.

\cite{Jahn20076289} suggested that an interplay between molecular
diffusion of alcohol and laminar convection results in the formation
of liposomes of controllable distribution.  The size control was
achieved in the range 50 to 150~nm by adjusting the ratio of flow
rates and the over all flow rate.  While the observations provided an
empirical way to achieve a control of the liposome size, we do not
have a quantitative model for predicting the sizes. This lack of
knowledge does not confine itself to the microfluidic method alone. In
the entire gamut of ethanol injection methods \citep{Batzri19731015,
  Kremer19773932, Wagner2002259, Ishii1995483} there is no theoretical
model yet that can explain the size selection or the formation of
multi-lamellar liposomes in some methods.

We hypothesise that the nature of flow and mixing has a significant
effect on determining the outcome of the self-assembly.  To show this,
we first choose a simpler configuration---core-annular two stream
flow, and compare the liposome formation in this system along with
other methods of mixing the two streams.  We find interesting
connections of the liposome size with the nature of micro-mixing.

An additional contribution in this paper is in the method of device
fabrication itself. While the channel geometry suggested in
\citet{Jahn20042674} is easy to replicate, it requires sophisticated
instruments to create the initial moulds.  The channel flow also poses
a theoretical difficulty in understanding the interaction of flow and
diffusion, as there are no simple analytical solutions to the velocity
and concentration profiles. A co-axial core-annular (or core-sheath)
flow is easily amenable to simple calculations owing to the axial
symmetry of the solution.  Here we introduce a simple method to
fabricate a co-axial hydrodynamic focussing device to synthesise
liposomes following the ideas of glass-PDMS hybrid microfluidic device
\citep{Jeong2004576}, and a method of swelling of PDMS in an organic
solvent to create channels \citep{Verma200610291}.  All of these can
be easily reproduced in any laboratory without requirement of
sophisticated instruments and environments to carry out
photo-lithography (such as, spin-coating, mask alignment tools, gold
room etc.).

We first outline the procedure to fabricate the co-axial
(core-annular) flow generating device in the Methods
section. Following this, we show various studies using the axial
flow-focussing method---the effect of flow rates, type of mixing,
lipid concentration, ethanol concentration, lipid chemistry on the
size of liposomes.  We then present evidence that supports our claim
that it is micromixing that leads to selection of larger size
liposomes formed within the micro channel.

\section{Materials and Methods}
\subsection{Materials}
The phospholipids used to synthesize the liposomes in the present
method are: 1,2-dimyristoyl-sn-glycero-3-phosphocholine (14:0-DMPC);
1,2-dipalmitoyl-sn-glycero-3-phosphocholine (16:0-DPPC) (Avanti Polar
Lipids), 1,2-dioleoyl-sn-glycero-3-phosphocholine (18:1-DOPC) (Sigma
Aldrich), and Hydrogenated Soy-phosphocholine (HSPC, gift sample from
Lipoid) is used as purchased without further purification.  Ethanol
(AR grade, Merck) is used to dissolve phospholipids. 

All the lipids from Avanti Polar Lipids were in a solution form with
chloroform (20 mg per ml) as solvent. An appropriate volume of lipid
solution is taken out in a round bottom flask (RBF) and chloroform is
evaporated from lipid solution using a rotary evaporator. This forms a
thin film on the glass surface of the RBF. The RBF is then kept in a
vacuum desiccator overnight to remove traces of chloroform. An
appropriate volume of ethanol is added to the RBF to make various
concentrations of the phospholipids in ethanol.

The aqueous phase used in all the experiments is Milli-Q water which
has been analysed by light scattering to ensure that no extraneous
particles are present.  The dye used for visualisation experiments is
a fountain pen ink (Camlin Royal Blue) dissolved in ethanol, which is
carefully filtered through a 0.2~$\mu$m filter paper, to limit the
size of particles.  After filtering it was found that the solid dye
particles had a unimodal distribution of sizes with a mean diameter
235$\pm$15~nm (as determined by Dynamic Light Scattering, Zetasizer,
Malvern, UK).

\subsection{Microfluidic device fabrication procedure}

\begin{figure*}[tb]
\begin{center}
\includegraphics[width=5.0in]{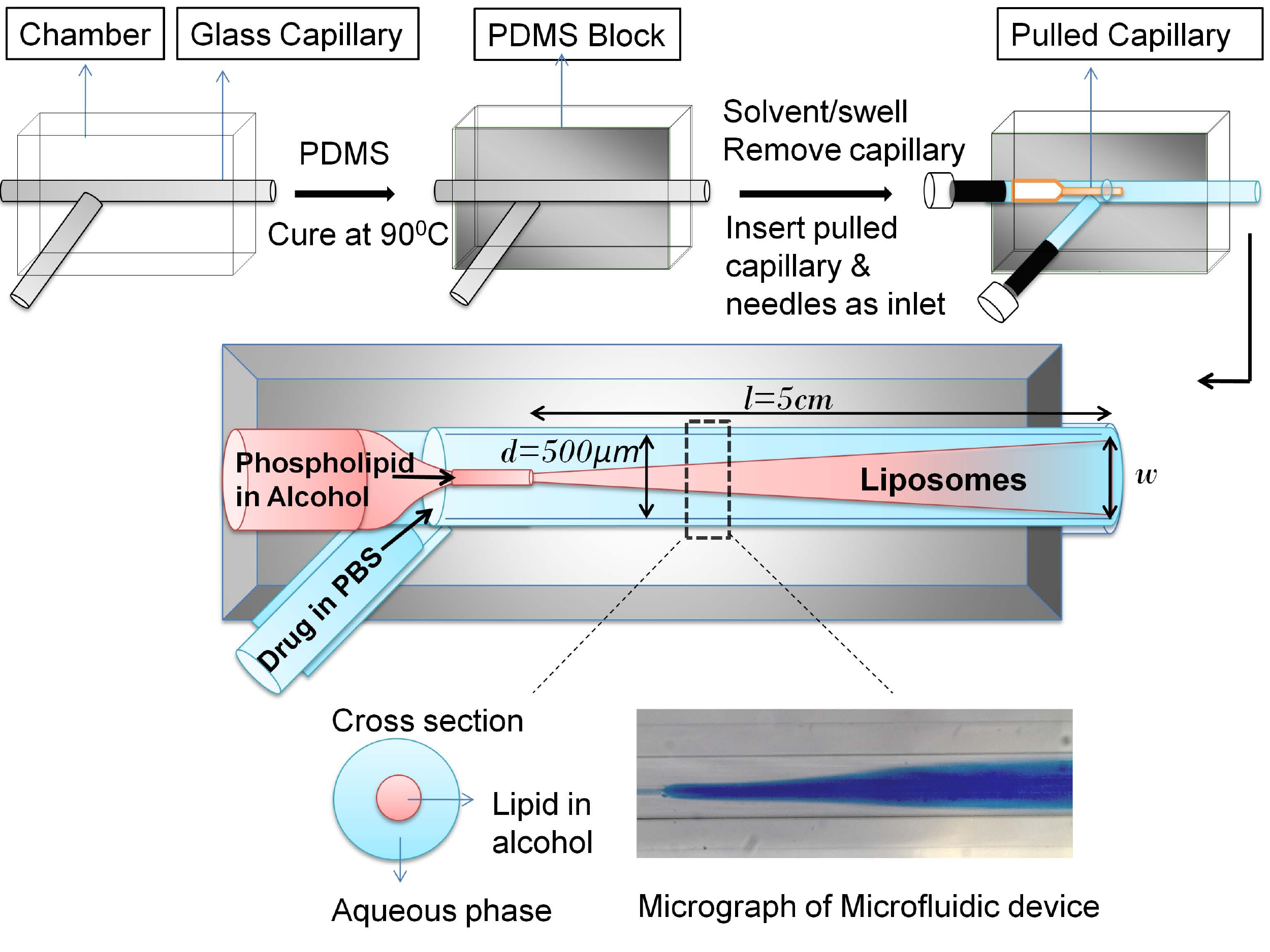}
\caption{ Top: Schematic diagram of core-annular microfluidic device
  fabrication sequence. Middle: Final assembled configuration of the
  entire setup. Bottom: Cross section of axial flow focusing and a
  sample photograph showing the spreading of the dye from the core
  stream into the annular region. The liposomes are expected to be
  formed in the mixed region.}
\label{fig:Devicedevlp}
\end{center}
\end{figure*}

Here we describe the method to prepare the co-axial flow microfluidic
device in some detail.  The method is easily reproducible in any
laboratory with basic chemicals and instruments. An open chamber is
prepared with the help of a glass slide and strips of double sided
tape as side walls. Two Borosilicate glass capillary tubes (30~mm and
90~mm in length, 1~mm outer diameter (OD) and 0.5~mm inner diameter
(ID): GD-1, Narishige, Japan) are joined together using using
Araldite\textsuperscript{\textregistered} (which is soluble in
chloroform) into a Y-shaped geometry as shown in
\Figref{fig:Devicedevlp}, and is inserted through a hole in the tape
inside the chamber. Liquid Polymer polydimethyl siloxane (PDMS) (Dow
corning Sylgard 184) along with 10\% crosslinker is then poured in the
chamber and left for curing at 90\degC\ for two hours.  The glass
slides are removed and the now solid PDMS block is left in chloroform
for two hours to allow it to swell (degree of swelling is
$>$150\%). The side glass capillary is now detachable, leaving a
channel in its place.  The main capillary is retracted partly up to
the Y-junction, and left inside the PDMS block to serve as a
hydrophilic side wall for the outlet stream.  A pulled capillary
(pulled to an ID of 40~$\mu$m, with a glass micropipette puller:
Narishige PC-10, Japan), is then inserted through the hollow PDMS
channel to go past the Y-junction through the main capillary. The
length of the pulled capillary is such that the entire tube is inside
the PDMS channel. Two blunt needles of appropriate size are then
inserted through inlet channels up to the glass capillary. The PDMS
block is then left at room temperature for the chloroform to evaporate
and regain its original shape and size, now tightly holding the
capillary and the needles. Since the flow usually has a small Reynolds
number, the flow through the side channel attains the fully developed
flow in a short distance, and can be used to hydrodynamically focus
the core flow. Teflon tubing is used to connect the microchannel to
the syringe \citep{Phapal2011}. Two Syringe pumps (NE1000, New Era
pumps Inc.) are used for pumping the fluids inside microchannel.

\subsection{Experimental setup for liposome synthesis}
The vesicles synthesis from the solution phase is only possible above
the phase transition temperature ($T_{\mathrm{m}}$) of the lipid
membrane, so it is necessary to maintain a temperature of the system
above this. When required, the microfluidic device along with the long
teflon tubings are kept inside an incubator with the temperature
controlled in the range 27\degC\ to 67\degC.  The experiments with
DMPC lipids were performed at 35\degC.

\subsection{Estimation of mean diameter and polydispersity of microfluidic liposomes}

The resultant liposomes are characterized by a single angle
back-scatter Dynamic Light Scattering (DLS) or Photon Correlation
Spectroscopy (PCS) (Zetasizer nano ZS, Malvern, UK).  He-Ne laser at
633~nm wavelength and 4 mW power is used and the scattered light is
detected at an angle of 173\degree. The size and size distribution is
calculated using software DTS 6.1 (supplied by Malvern). The
Polydispersity and the average liposome diameter are evaluated using
cumulant analysis method and the weight-weight distribution of diameter
is evaluated using multimodal distribution analysis. Three independent
experiments are carried out for liposome synthesis experiments and for
each of them, three measurement runs are performed).  Ethanol
present in the dispersion has a large influence on the viscosity of
solution and a negligible effect on the refractive index; these
corrections are made using the DTS 6.1 software (Malvern) and were found
to be in good agrement with reported values \citep{Song20081367}.

\subsection{Estimation of size and lamellarity of liposomes}
Transmission Electron Microscopy (TEM) (JEM-2100F, Jeol) is performed
to check the lamellarity of the liposomes. An electron dense material,
2\%w/v phosphotungstic acid (PTA) (SD-Fine chemicals India), is used
for negative staining . A drop of the liposome solution (10~$\mu$l) is
placed carefully on the TEM grid with the help of a micropipette and
kept for air drying for 10 min, followed by placing a drop of negative
stain. The sample is air dried again before observing at an operating
voltage of 120~kV.

Environmental Scanning Electron Microscopy (ESEM) (Quanta-250 FEG,
FEI) analysis was performed using a Wet-STEM-detector and the
dispersion is observed at a temperature of 4\degC\, 100\% humidity,
20~kV accelerating voltage (0.14~nA current), and 840~Pa pressure. Wet
STEM allows observation of liposomes keeping its environment fully
hydrated i.e. in its native state. The protocol followed is described
in the literature \citep{Mohammedetal2004,Barbaraetal2011}.

\section{Results}

In the sections below we first report the observations of various
influences on the size and nature of liposomes formed.  The co-axial
(core-annular) microfluidic device is used for all these experiments,
unless otherwise stated.  Lipid solution in ethanol is used as the
core fluid stream and water is used in the outer stream.  The error
bars in the plots denote the standard deviation about the mean.

\subsection{Effect of flow on the mean liposome size}

The ratio of the volumetric flow rates of the aqueous (annular) and
lipids dissolved in ethanol (core) has been shown to have a
significant effect on the size of liposomes in the rectangular cross
section geometry \citep{Jahn20076289}.  We confirm that a similar
behaviour is reproducible even in the axial flow focussing geometry.
In a later section we also provide a plausible explanation for this
behaviour.

\begin{figure}[tb]
  \begin{center}
\includegraphics[width=0.9\linewidth]{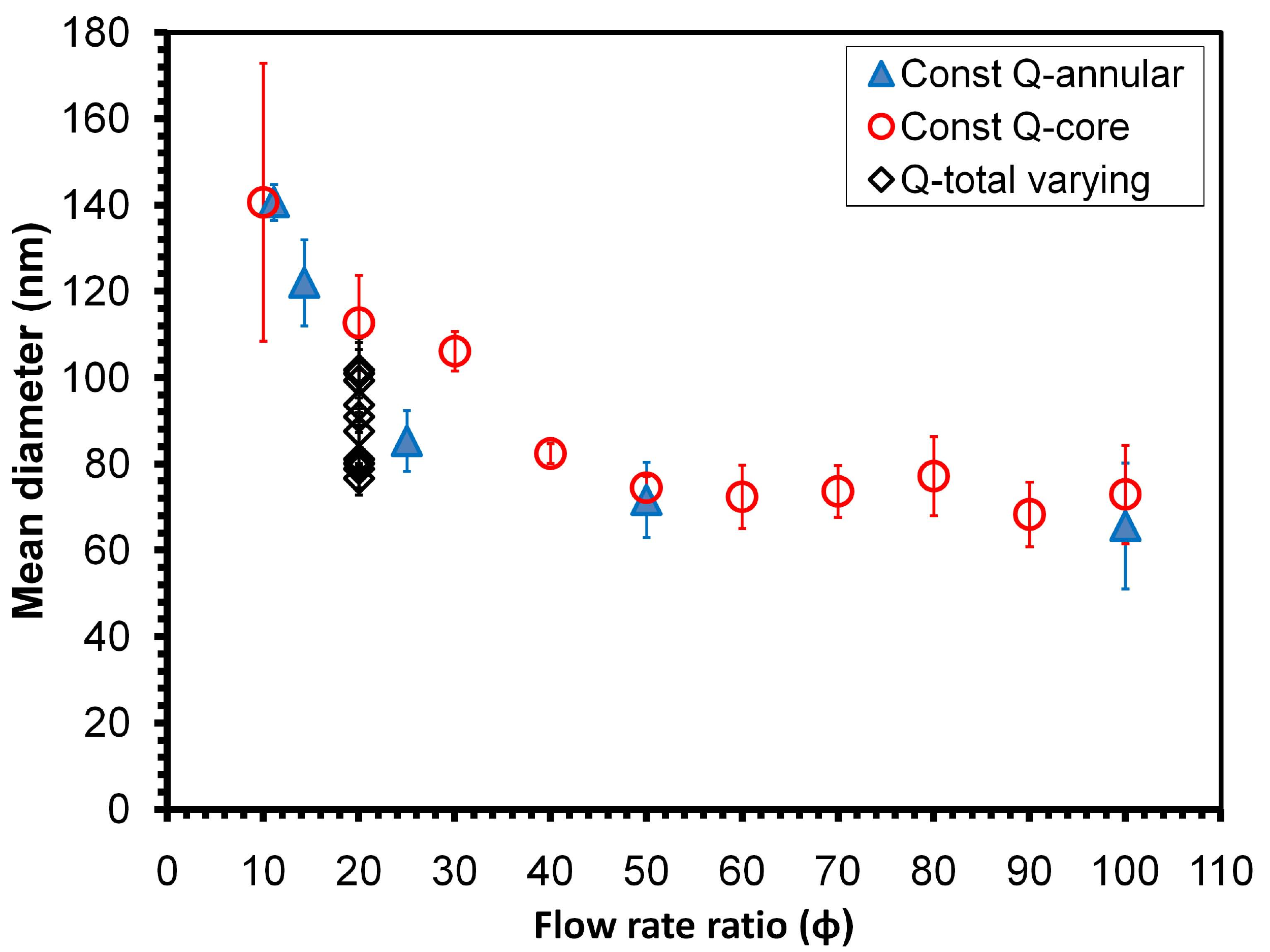}
\caption{Effect of the flow rate ratio $\phi$ (outer to inner) and
  total flow rate on the size of liposome formed. The mean diameter is
  obtained by DLS. Triangles: $Q_{\mathrm{W}}$ (aqueous in the annular
  region) is kept constant at 100~$\mu$l/min and $Q_{\mathrm{E}}$
  (lipidic ethanol in the core region) increases going from right to
  left. Circles: $Q_{\mathrm{E}}$ is constant at 10~$\mu$l/min and
  $Q_{\mathrm{W}}$ increases from 100 to 1000~$\mu$l/min (going from
  left to right). Diamonds: $\phi = 20$ is kept constant but the total
  flow rate $Q$ is varied from 21 to 945~$\mu$l/min, by increasing
  both the flow rates. }
\label{fig:phivar}
\end{center}
\end{figure}

The flow rate ratio is denoted here as $\phi$, and defined as
\begin{equation}
  \phi = \frac{Q_{\mathrm{annular}}}{Q_{\mathrm{core}}} =
  \frac{Q_{\mathrm{W}}}{Q_{\mathrm{E}}} 
\end{equation}
where $Q_{\mathrm{annular}}$ (or $Q_{\mathrm{W}}$ ) is the outer
(water) volumetric flow rate and $Q_{\mathrm{core}}$ (or
$Q_{\mathrm{E}}$) is the inner (ethanol) volumetric flow rate.  It is
seen from \Figref{fig:phivar} that at lower $\phi$, the size decreases
with increase in $\phi$, and plateaus to a nearly constant value at
large $\phi$.  In this study $\phi$ has been varied by two ways, one
by keeping the inner flow rate constant and other by keeping the outer
flow rate constant.  The liposome sizes have been determined by PCS
(single angle DLS).

It is observed that in both the cases (constant $Q_{\mathrm{W}}$
and constant ${Q_{\mathrm{E}}}$), the qualitative behaviour is the
same--a decrease followed by a plateau. The large $\phi$ plateau for
the liposome sizes has not been reported in earlier works of
microfluidic synthesis \citep{Jahn20076289,Jahn20102077}, probably
owing to the limitation of a channel geometry where obtaining a stable
flow is difficult at high flow rate ratios.  Though not shown here,
in the axial flow focussing we can obtain flow rate ratios as high as
$\phi$= 500 (${Q_{\mathrm{E}}}$=5~$\mu$l/min and
${Q_{\mathrm{W}}}$=2500~$\mu$l/min).

Whereas the large $\phi$ plateau of the liposome size is nearly equal
in both the cases, there seems to be a difference in the size of
liposomes between the two cases for intermediate flow rate ratio
around $\phi \approx 20$. This suggests that there could be
other parameters influencing the size of liposomes, rather than just
the flow-rate ratio, as thought previously \citep{Jahn20076289}.

\begin{figure}[tb]
\begin{center}
\includegraphics[width=0.9\linewidth]{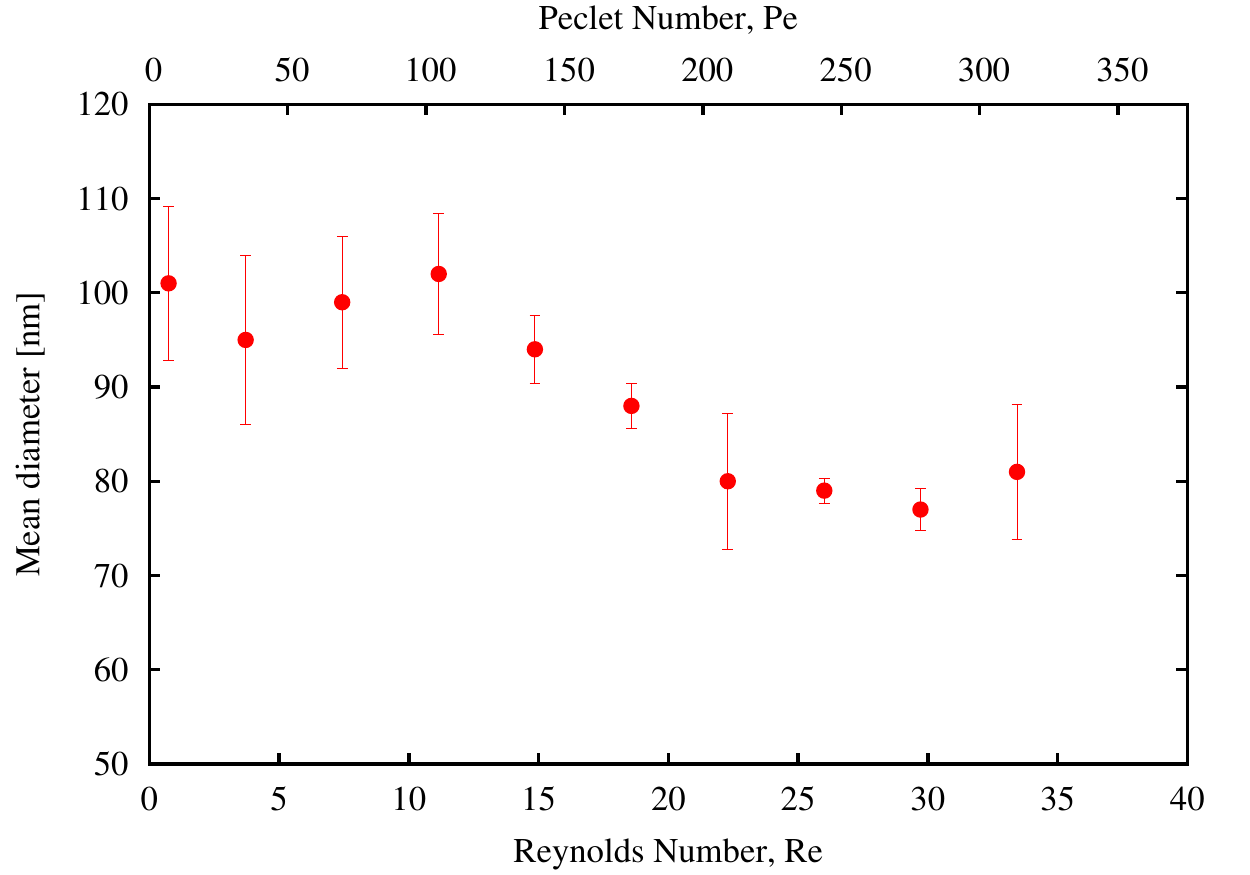}
\caption{Effect of the total flow rate on the size of liposomes. The
  Reynolds number Re is shown as one measure of the total flow rate.
  Also shown is an equivalent axis in terms of the Peclet number \Pe.
  The experiments are conducted at a constant $\phi = 20$, and varying
  total flow rate $Q$ such that $Q_{\mathrm{E}} =
  5,10,\ldots,45,50$~$\mu$l/min.}
\label{fig:Qtot}
\end{center}
\end{figure}

The presence of other influences on the size is more clearly seen when
the flow rate ratio $\phi$ is kept constant, while increasing the
overall flow rate $Q = {Q_{\mathrm{W}}} + {Q_{\mathrm{E}}} $.  We show
this in \Figref{fig:Qtot}, where the size of liposomes, at a constant
flow rate ratio $\phi = 20$, is plotted against the Reynolds number
$\Rey = d \, U\, \rho/\mu$, where $U$ is the superficial velocity $U
\equiv 4 Q/\pi\,d^{2}$, $d$ is the inner diameter of the outer
channel, $\rho$ is the mean mass density and $\mu$ is the viscosity of
the fluid (here taken as that of water). At low velocities, the size
is nearly constant, at a value around 100~nm, comparable to the low
$\phi$ sizes shown in \Figref{fig:phivar}, whereas at higher
velocities it approaches a value close to 80~nm, close to the large
$\phi$ plateau of \Figref{fig:phivar} (We can confidently reject the
null hypothesis that the mean size of liposomes for the group $\Rey
\lesssim 15$ is the same as that for the group $\Rey > 20$,
with a $p$-value less than $2 \times 10^{-8}$ determined by the Welch
$t$-test). This variation is also shown in \Figref{fig:phivar} (as
diamonds at constant $\phi=20$).  Again this observation has not been
reported earlier, possibly because in the regimes explored in
\cite{Jahn20076289} the overall flow rate was incidentally varied only
in the region where it had no influence.

The flow rate ratio $\phi$ also specifies the ternary component
composition (water:ethanol:lipid) in the final collected sample. A
lower $\phi$ implies a higher fraction of lipid and ethanol in the
ultimate mixture. The size of the liposomes cannot, therefore, be
characterised in a simple ternary component diagram
~\citep{Ishii1995483} which are commonly used for equilibrium systems.

Increasing the overall flow rate amounts to two obvious influences on
the local transport phenomena: increase in the local shear rate and an
increase in the Peclet number (that measures the relative importance
of axial convection to the radial diffusion of species) defined in
this case as:
\begin{equation}
  \label{eq:Pe}
  \Pe=\frac{U \, d^{2}}{\mathcal{D}\, L}
\end{equation}
where, $\mathcal{D}$ is the diffusivity of a solute and $L$ is a
characteristic length along the axial direction.  The solute in this
case is ethanol, which dissolves and diffuses from the core region
into water in the annular region.

The influence of shear rate on the dynamics of self assembly of lipids
into disc-like micelles, and further aggregation of these micelles
towards larger domain growth and closure to form lipid vesicles, can
be one possible influence.  However, to the best of our knowledge this
has not been studied, and therefore, we summarily leave out exploring
this possibility.

\Figref{fig:Qtot} indicates that increase in the Peclet number \Pe\
(only $U$ is varied in \Eqref{eq:Pe}) leads to a smaller sized
vesicle.  The value of \Pe\ for this study is varied from \OrderOf{10}
to \OrderOf{100}. A large $\Pe \gg 1$ implies that the two streams
hardly mix by diffusion inside the channel of the device. Mixing of
the two streams is essential for the formation liposomes. It is only
when water and ethanol inter-diffuse and the lipids can no longer
remain in the dispersed form (due to the energetic costs in the
presence surrounding water molecules) that they self-assemble to form
vesicles. \Figref{fig:Qtot} indicates that \Pe\ values are very large
for the entire range of flow rates studied, and there is a variation
in the liposome sizes. This is counter intuitive.  $\Pe \gg 1$ implies
the flow inside the channel does not lead to (diffusive) mixing of the
streams and therefore does not induce any self assembly. The mixing of
the two streams takes place outside the channel as the outlet flow is
collected in a cuvette, in which case the liposome size should have
been uniform for all $\Pe \gg 1$.  This suggests that there could be
other phenomena occurring in the channel at $\Pe \approx 200$ that is
inducing an increase in the liposome size.

\subsection{Influence of type of mixing}
\begin{figure*}[tb]
\begin{center}
\includegraphics[width=0.9\linewidth]{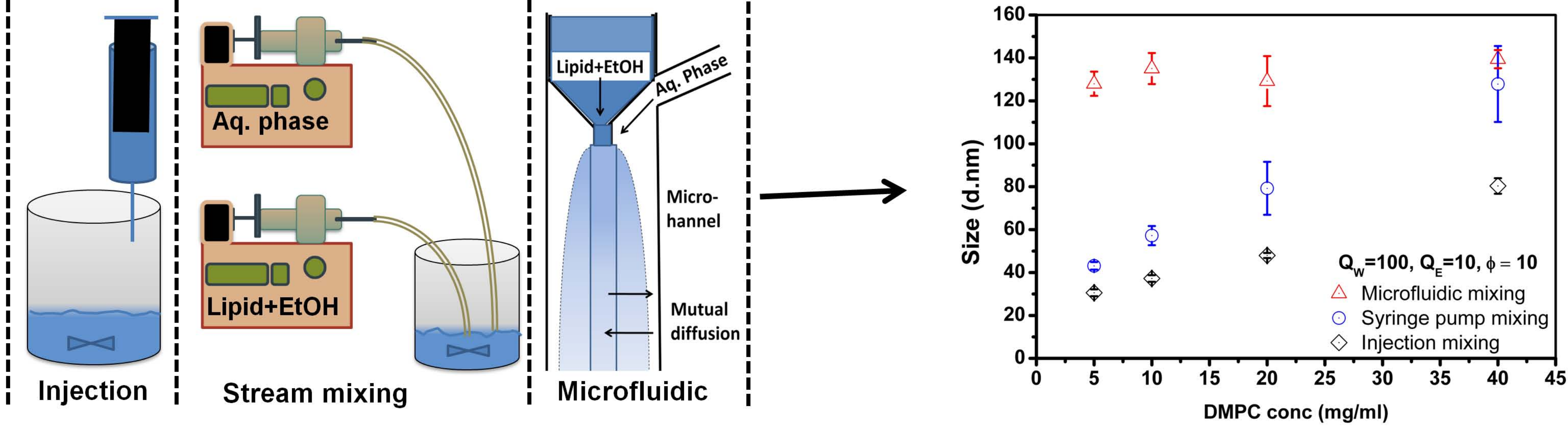}
\caption{ Three types of mixing methods employed to study the effects
  of bulk mixing: (a) Bulk Injection---by injecting lipid-ethanol in
  the aqueous phase under continuous stirring; (b) Stream mixing
  (Syringe Pump)---mixing of lipid-ethanol phase stream with an
  aqueous phase stream each driven the help of syringe pumps and fine
  Teflon tubing. The flow rate of lipid solution was 10~$\mu$l per min
  and aqueous phase was kept at 100~$\mu$l per min. (c) Laminar
  Microfluidic mixing. The flow rates were same as in stream mixing.
  In all the above methods, the volumetric ratio of ethanol to aqueous
  phase is 1:10. The plot on the right hand side shows the liposome
  size as a function of concentration of lipids and the type of
  mixing.}

\label{fig:different-mixing}
\end{center}
\end{figure*}

Though several variants of ethanol injection methods have been
employed \citep{Batzri19731015, Kremer19773932, Wagner2002259,
  Ishii1995483}, there is no reported analysis of what influences the
size of liposomes.  Here we employed various types of mixing the
fluids and observed the dependence of the liposome size.
\Figref{fig:different-mixing} shows that the liposome size is a
function of lipid concentration for various ways of mixing the
lipid/ethanol with the aqueous phase: (a) mixing of lipid-ethanol by
injection in the aqueous phase under continuous stirring, maintaining
the ratio of volumes at 1:10 (equivalent to the flow situation of
$\phi =10$) (b) bulk vigorous mixing of lipid-ethanol phase stream
with an aqueous phase stream, each of them pumped using calibrated
syringe pumps. Here the flow rate of lipid solution was 10~$\mu$l/min
and aqueous phase was 100~$\mu$l/min, and (c) Mixing by microfluidic
device, here again the flow rates were similar as in (b). In each
case, the ratios of lipid:ethanol:water are kept constant. It is clear
from \Figref{fig:different-mixing}, that direct mixing by turbulence
yields smaller vesicles (SUVs), mixing in streams gives slightly
larger vesicles and microfluidic mixing gives even larger
vesicles. The first two methods showed a dependence of size on the
lipid concentration in ethanol but in the case of microfluidic mixing,
liposome size seems independent of lipid concentration, for even as
high as 80~mg/ml of DMPC. Methods (a) and (b) also show polydispersed
population (not shown here).

Though the implication of the effect of concentration is not evident,
we can conclude on the effect of nature of mixing on the size of
liposomes.  The stronger the nature of mixing the smaller the size.
In order to form larger liposomes, the bilayer membrane discs has to
grow to large sizes before closing upon themselves.  However, this
growth seems to be hindered by the stronger flow, leading to earlier
closure and therefore smaller liposomes.

\subsection{Liposome characterisation}

\begin{figure}[tb]
\centering
\includegraphics[width=0.9\linewidth]{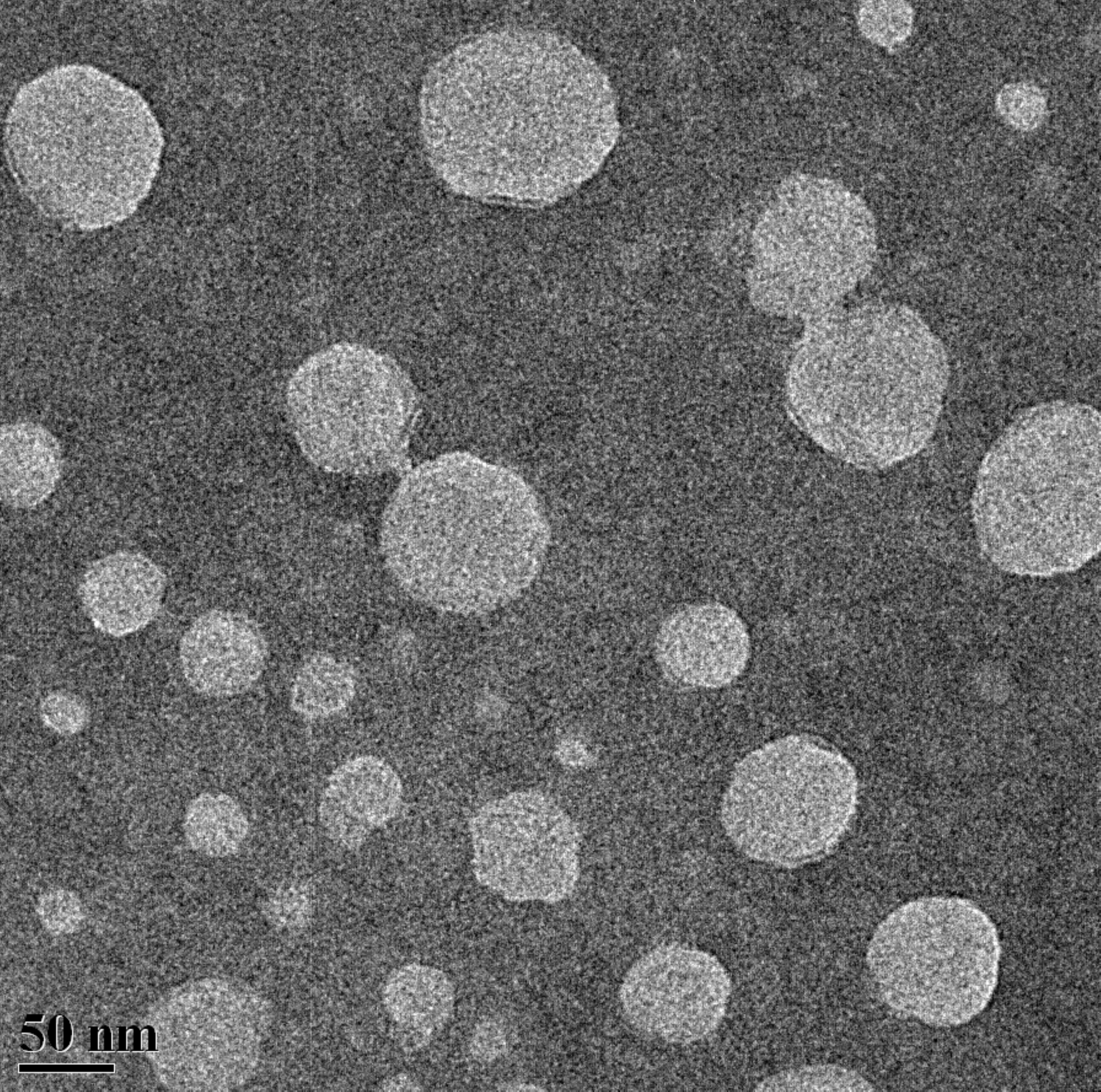}
\caption{TEM micrographs of DMPC liposomes showing unilamellar
  membrane liposomes. Otherwise, TEM with a contrast providing agent
  will display striations of a multi-lamellar structure
  \citep{Leeetal2013}. The synthesis conditions employed here are:
  10~mg/ml DMPC, 10~$\mu$l/min ethanol flow rate and 200~$\mu$l/min
  water flow rate.}
\label{fig:TEM-MF}
\end{figure}

\begin{figure*}[tb]
\centering
\includegraphics[width=0.45\linewidth]{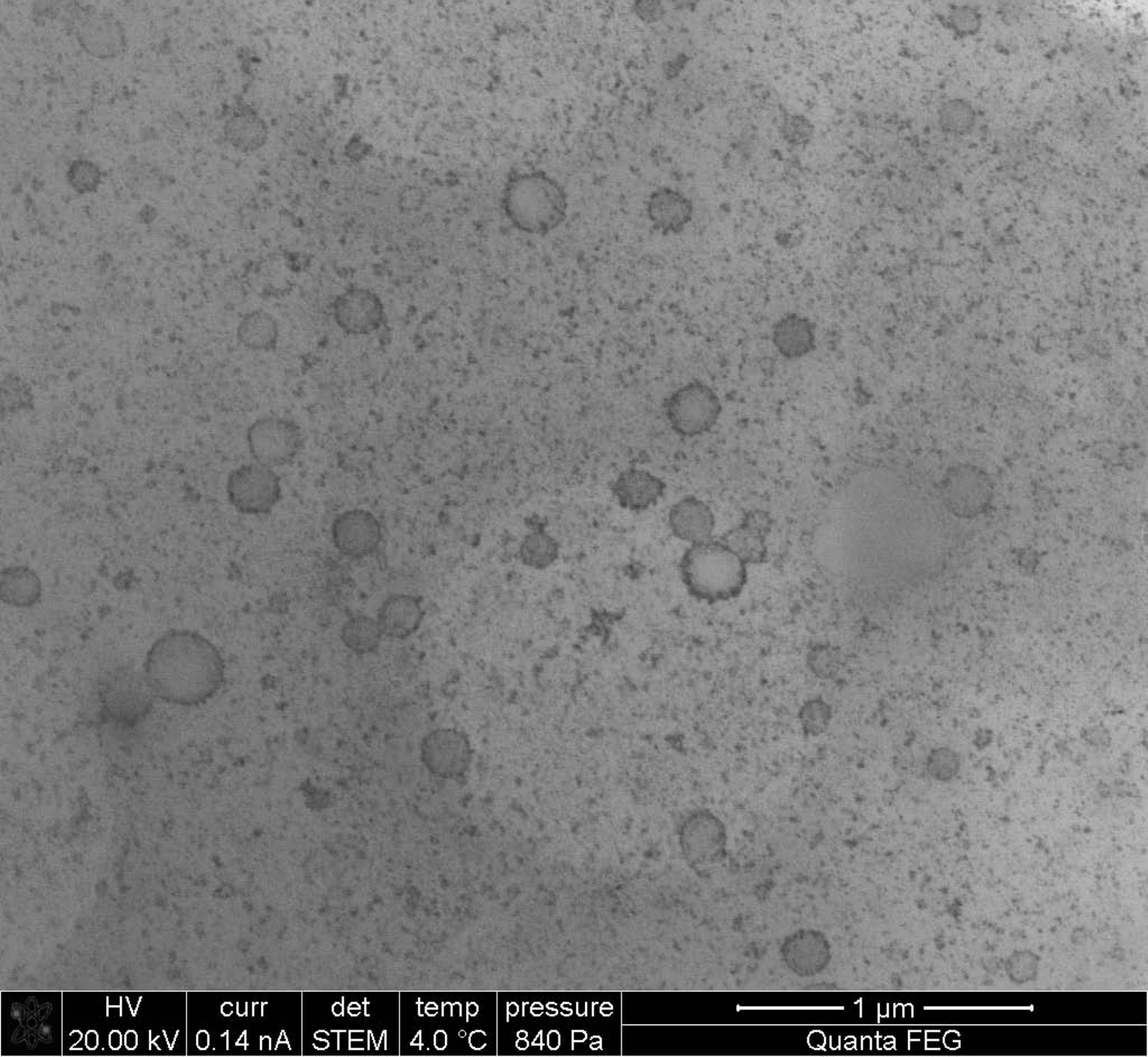}
\includegraphics[width=0.45\linewidth]{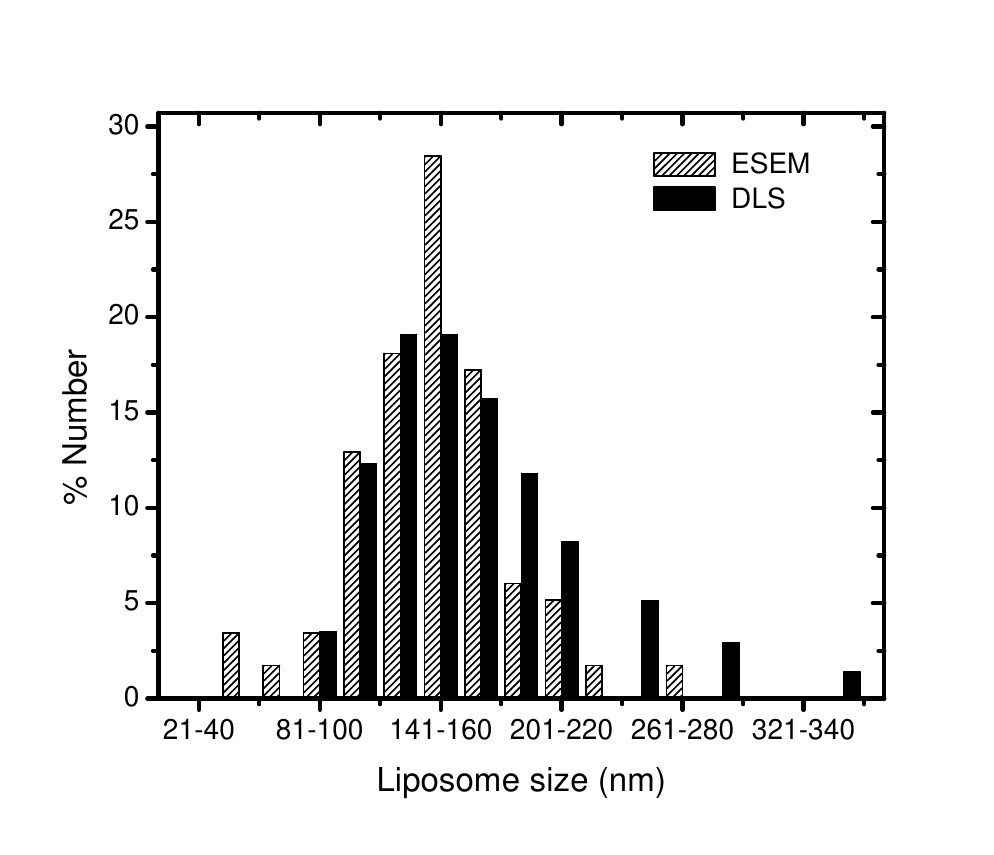}
\caption[ESEM]{An Environmental Scanning Electron Microscopy (ESEM)
  micrograph of HSPC liposomes obtained with
  $Q_{\mathrm{E}}$=10~$\mu$l/min, $Q_{\mathrm{W}}$=400~$\mu$l/min,
  $\phi$=40. Right: Comparison of number distributions of liposome
  size obtained by ESEM and DLS. For ESEM, the number distribution was
  obtained from 97 liposomes in four micrograph images.}
\label{fig:ESEM}
\end{figure*}

The liposomes formed by the co-axial flow focussing device have also
been characterised by the usual techniques. The particle size
distribution by DLS of liposomes is nearly mono-disperse with the
polydispersity index is around 0.2$\pm$0.05.  \Figref{fig:TEM-MF}
shows a Transmission Electron Microscope (TEM) micrograph of DMPC
liposomes, clearly displaying the unilamellar bilayer membranes using
the negative staining method used for visualizing.  \Figref{fig:ESEM}
shows the intact and spherical shaped liposome, fairly uniform size
distribution of liposome prepared with microfluidic
synthesis. \Figref{fig:ESEM} also shows a comparison of particle size
distribution obtained with DLS along with that obtained by visual
analysis of ESEM results, indicating that the size distribution
obtained in DLS is a reliable measure of the actual distribution.

\subsection{Effect of lipid concentration}
\begin{figure}[tb]
\begin{center}
\includegraphics[width=0.9\linewidth]{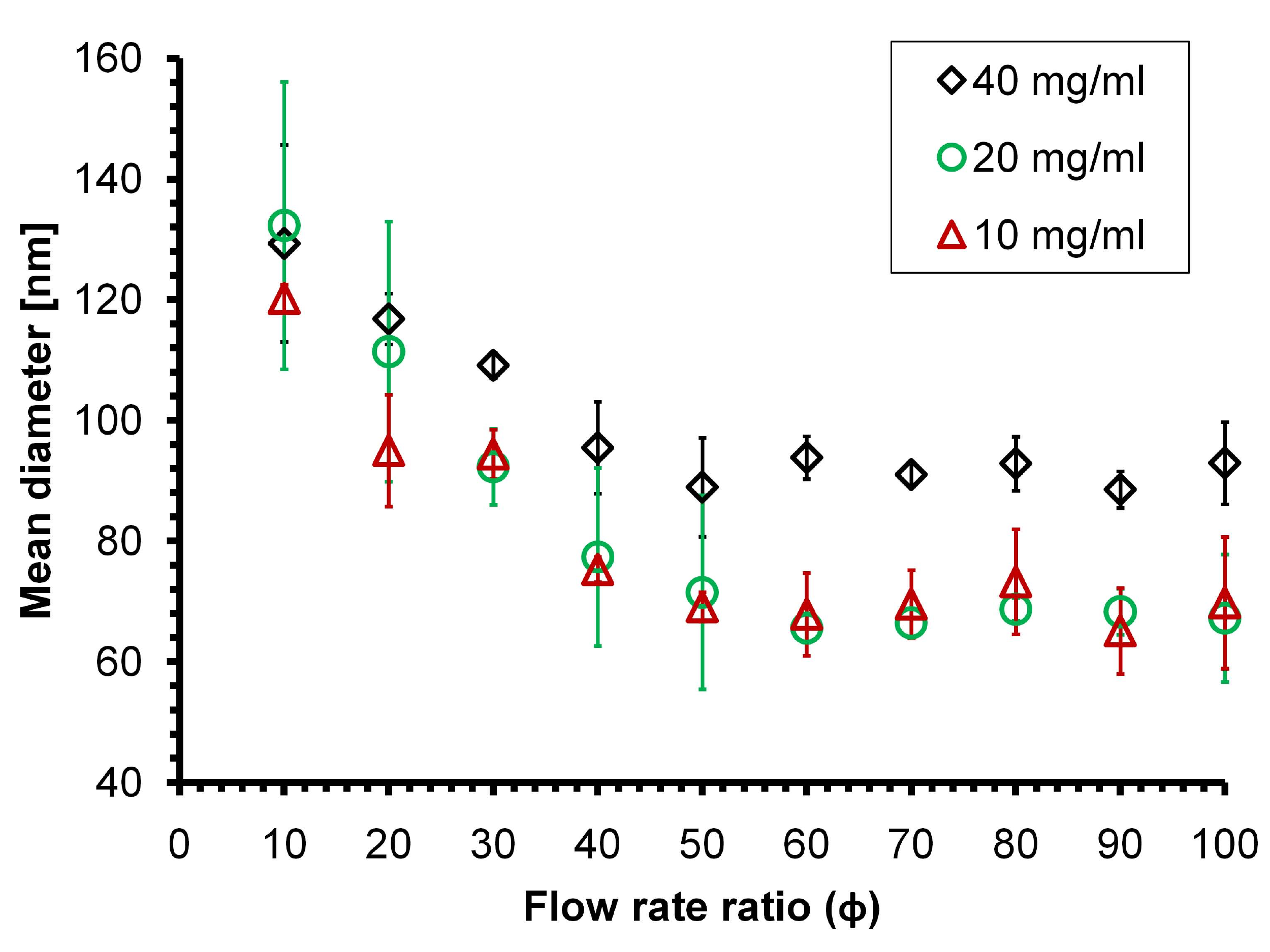}
\caption{Effect of phospholipid concentration (in ethanol) on the
  liposome size. At lower $\phi$, size is insensitive to the
  concentration whereas at higher $\phi$, the higher concentration of
  40~mg/ml produces larger sized liposomes.}
\label{fig:conceffect}
\end{center}
\end{figure}

In coacervation techniques, it is well known that liposome size and
polydispersity increases with lipid concentration
\citep{Kremer19773932, Wagner2006311}. We report the effect of three
different DMPC stock concentration--10, 20 and 40 mg/ml of ethanol
using the axial flow-focussing device (\Figref{fig:conceffect}). For
10 and 20~mg/ml of lipids, the concentration does not seem to affect
the size of liposomes. The concentration of lipids seems to have an
effect only at large values of $\phi$. At smaller values of $\phi$ the
concentration of lipids does not significantly influence the size of
the liposomes. This also happens to be the region of $\phi$ where we
see a variation in the size upon changes in $\phi$. This again
indicates that there are other phenomenon that influences self
assembly in this region of flow.

\subsection{Effect of membrane composition}
\begin{figure}[tb]
\begin{center}
\includegraphics[width=0.9\linewidth]{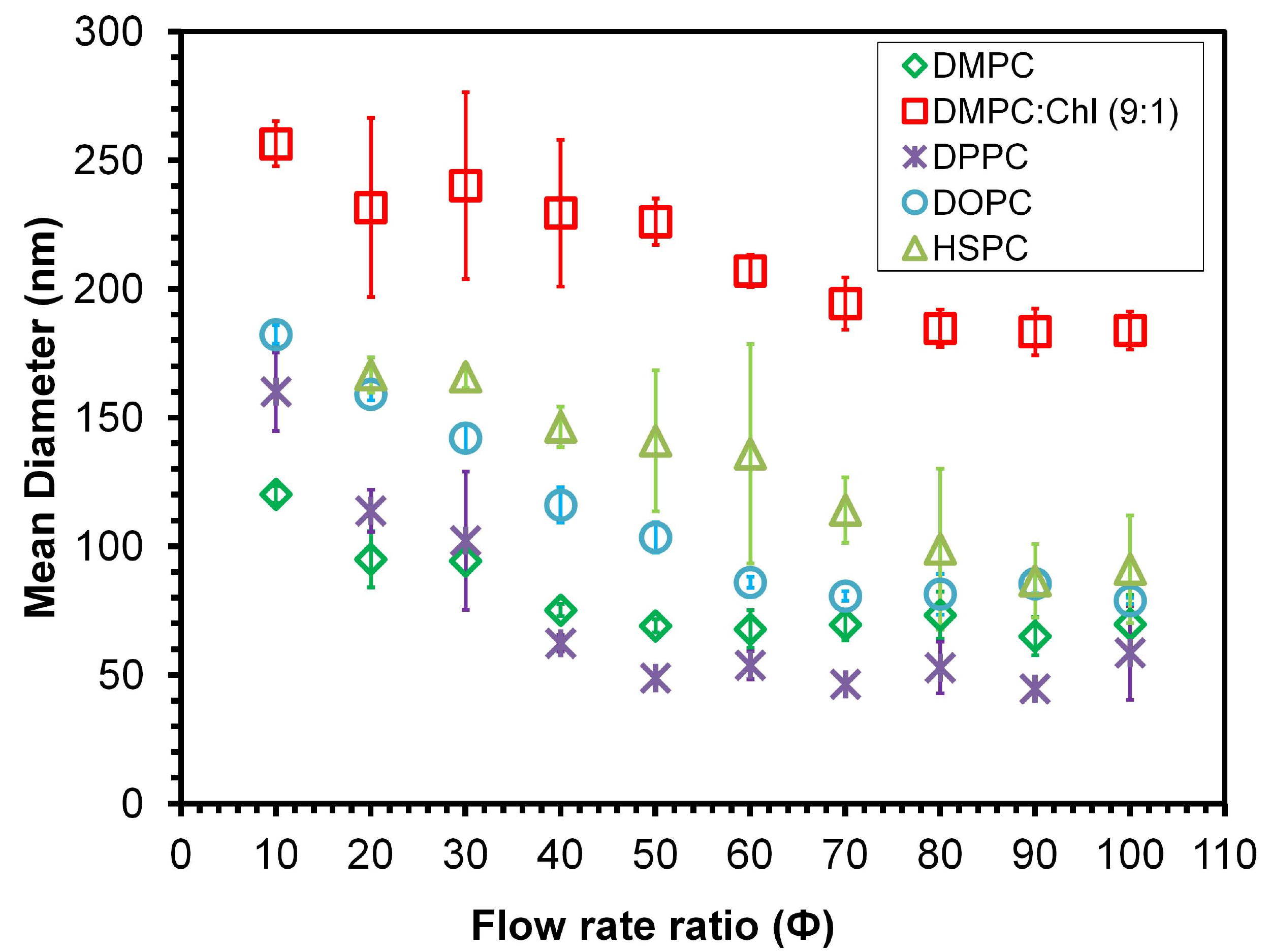}
\caption{Liposome size as a function of flow rate ratio $\phi$ and
  membrane composition.}
\label{fig:all_lipids}
\end{center}
\end{figure}

The mean diameter of vesicles is also a function of phospholipid
chemistry and membrane composition as seen from
\Figref{fig:all_lipids}.  To study the effect of cholesterol on
liposome size, DMPC:Cholesterol in a ratio 9:1~\%w/w is dissolved in
ethanol, so that the final lipid concentration in the stock solution
is 10~mg/ml.  The experiment was conducted at 35\degC\ inside an
incubator with controlled temperature. We see that addition of
cholesterol increases the liposome size.  Addition of cholesterol
increases the bending rigidity modulus $\Kb$ by about 50\%
\citep{Meleard19972616,Hofsasz20032192}.  The rigid bilayer disk
curves into vesicles by thermal fluctuations only when they have grown
in size above a critical dimension \citep{Helfrich1973693,
  Leng20031624}.  Increased stiffness implies the discs need to grow
larger before thermal fluctuation can induce a closure.

Secondly, we alter the lipid chain length and saturation by
considering other lipids in place of DMPC as shown in
\Tabref{tab:lipidchemistry}. DOPC (carried out at a room temperature
of 27\degC) produces vesicles larger in size than the DMPC lipids. The
bending modulus for DOPC (above transition temperature) is also about
50\% higher than DMPC lipids~\citep{Rawicz2000328}.  Liposomes with
DPPC (performed at 55\degC\ inside the incubator) shows a mixed
behaviour in comparison with DMPC lipids: larger at lower $\phi$ and
comparable at larger $\phi$. Owing to a larger chain length, it is
expected that DPPC has a larger bending modulus compared to DMPC.
However, both optical \citep{Leeetal2001} and neutron spin echo
measurements \citep{Setoetal2008} show that the bending modulus of
DPPC is not significantly distinguishable from that of DMPC.  Lastly,
the hydrogenated Soy-PC (HSPC) lipids is a mixture predominantly
composed of 18:0 acyl chains. The bending modulus of the membrane
formed from this mixture is not reported to the best of our knowledge,
however it should be higher than DMPC owing to the chain lengths. The
liposome size also obeys this trend.

In drawing the above conclusions we have intentionally avoided using
absolute values of the bending modulus as they differ significantly,
for the same lipid membrane, depending on the method used. We have
cited only those references where the same method was employed to
measure the modulus of two different lipids, thereby permitting a
relative comparison.

\begin{table*}[tb]
\begin{center}
  \caption{Number of acyl carbon atoms--chain length ($C$), Number of
    unsaturated bonds ($D$) per acyl group, Transition temperature
    ($T_{\mathrm{m}}$), and Molecular weight for various lipids
    employed in this work.}
\label{tab:lipidchemistry}
\begin{tabular}{|p{1.5cm}|p{2.8cm}|p{1.0cm}|p{1.0cm}|p{1.5cm}|}
\hline
\multicolumn{1}{|p{0.08\textwidth}|}{{ Name}}	&
\multicolumn{1}{p{0.24\textwidth}|}{{ $C$}} &
\multicolumn{1}{p{0.08\textwidth}|}{{ $D$}} &
\multicolumn{1}{p{0.1\textwidth}|}{{ $T_{\mathrm{m}}$} (\degC)}&
\multicolumn{1}{p{0.12\textwidth}|}{{ Mol Wt} (g/mol)}\\\hline
DMPC	& 14        & 0 & 23 & 678\\
DPPC	& 16        & 0 &41 & 734\\
DOPC	& 18        & 1 &-20 & 786 \\
HSPC	& 16~(11\%), 18~(89\%)& 0 & 55 & 784\\\hline
\end{tabular}
\end{center}
\end{table*}

\clearpage
\section{Discussion}

\subsection{Hypothesis for liposome size variation}
The results of the previous section can be summarised as follows (i)
there are two regimes of liposome formation: one in which the size
depends on the flow parameters but does not depend on the
concentration of lipids, and other in which the size is not dependent
on flow parameters but is influenced by concentration of lipids.
(ii) The dependence on the lipid chemistry and membrane composition is
in line with what is expected from a general understanding of the
bilayer growth and closure to for vesicles. Therefore, it is the
former observation that needs an explanation.

We hypothesise that the two regimes correspond to (a) Convective
instability mixing: Instability induced convective mixing within the
channel, and (b) Droplet mixing: Mixing of the streams outside the
channel when drops of the outlet are collected in a cuvette.
Core-annular flows of two miscible fluids where there is a viscosity
stratification leads to an instability at large Peclet numbers ($\Pe
\gg 1$) and small Reynolds numbers ($\Rey \equiv d\,U\,\rho/\mu$).
This was numerically shown by \cite{RangRama01} for channel flows and
by \cite{Selvametal07} for core-annular flows.  In the present case,
there is a viscosity variation in a non-monotonic manner, unlike that
considered in \cite{Selvametal07} where the viscosity variation is
monotonic.  \cite{DOlceetal08} have observed certain instabilities
experimentally which show pearl and mushroom-like appearance when
there is a monotonic increase in the viscosity from the core region to
the annular region. To the best of our knowledge such instabilities
have not been reported in the water-ethanol system.  Nevertheless, we
may expect a similar behaviour.

The reason for expecting the droplet mixing regime is that when the
\Pe\ number is large and when there is no convective instability, the
streams will only mix at the outlet.  This is supported by additional
facts that the dependence on the concentration of lipids is similar to
that seen when two streams are mixed in the various types of ethanol
injection methods shown in \Figref{fig:different-mixing}.

At present we do not have a theory to show the possible existence of
either of the above claims for the present system. However, it is
possible to test the consequences by carrying out visualisation
experiments on the same system.  In the following we report results of
experiments carried out in a similar system---the lipids in ethanol
being replaced by a dye---which us allows to visualise the nature of
flow inside the device.

\subsection{Flow instability due to viscosity tratification}
Core-annular instability is known to occur in immiscible flows. It was
shown in \cite{Selvametal07} a similar instability also occurs in
miscible fluids, and presence of a small amounts of diffusivity ($\Pe
\gg 1$) enhances this instability, particularly at small Reynolds
Numbers. Though the model is not identical to the present system, we
draw similarities in the nature of flow (core-annular) and the
presence of viscosity stratification.  In the present case $ 2< \Rey
<42 $ and $ 16< \Pe <507 $, based on water-ethanol diffusivity of
$\mathcal{D} = 1.28\times 10^-9$ ~\citep{Zhang2006}.

\begin{figure}[tb]
\begin{center}
\includegraphics[width=0.9\linewidth]{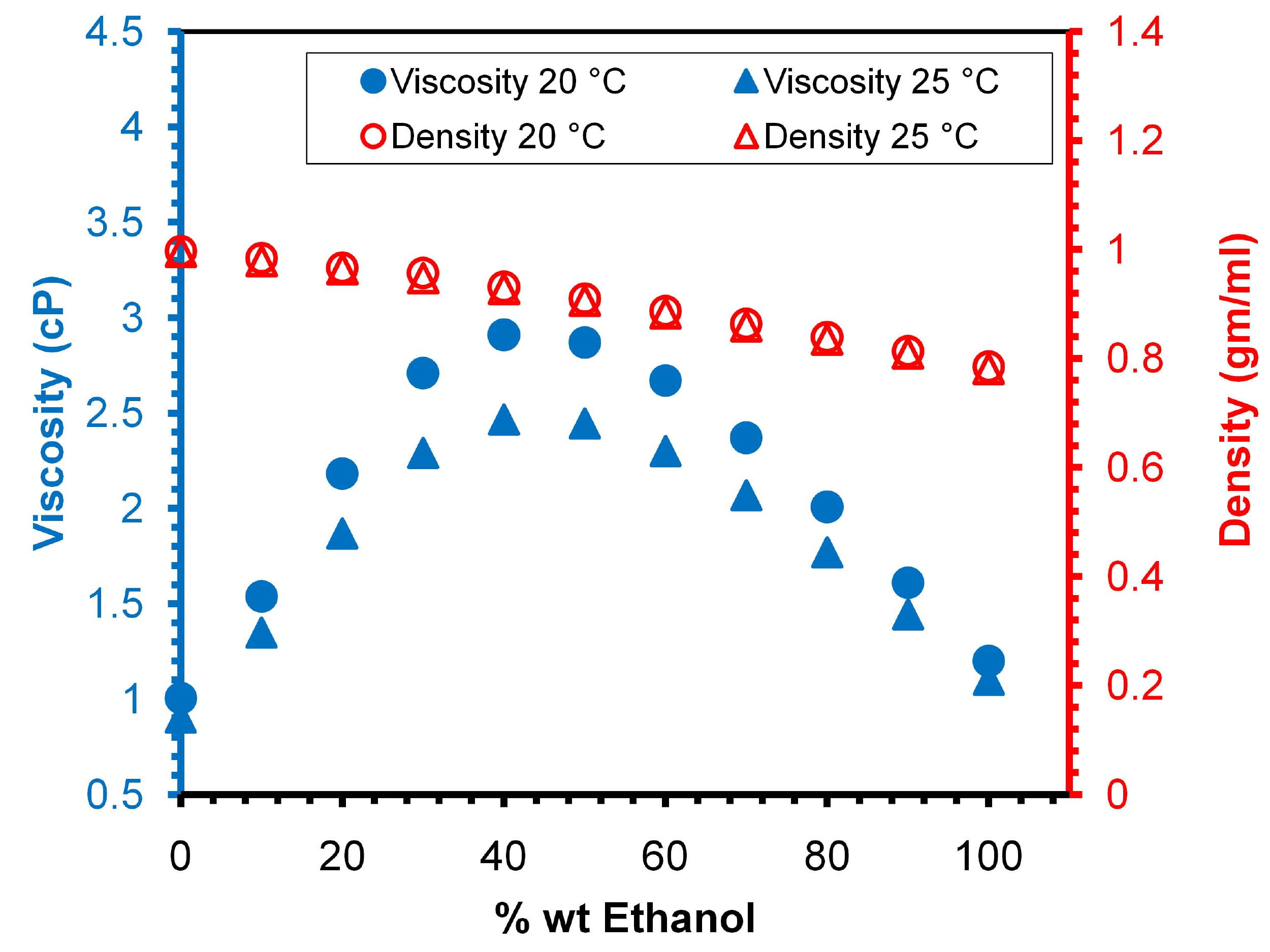}
\caption{Viscosity~\citep{Song20081367} and density~\citep{Perry1997}
  variation of ethanol--water system at different temperatures.}
\label{fig:weviscrho}
\end{center}
\end{figure}

The variation of viscosity of water-ethanol system
\citep{Song20081367} is shown in \Figref{fig:weviscrho}.  The maximum
viscosity of water-ethanol mixture at 20\degC\ is 2.68 mPa.S at 0.7
mole fraction of water; which is around three times the viscosity of
water.  \cite{DOlceetal08} observed pearl and mushroom like patterns
of the instability in the case of a water-natrosol system where the
core fluid was of a lower viscosity, and the viscosity increases
monotonically to about 25 times the core fluid
viscosity. \cite{Selvametal07} also reported instabilities to be
present in the opposite case of viscosity decreasing from the core to
the annular region. In the present case, we have both the
situations. The viscosity is maximum in a small mixed region in
between (at about 40\%w/w of ethanol), and decreases to about the same
value (1~cp) towards core and the periphery.

\begin{figure}[tb]
\centering
\includegraphics[width=0.9\linewidth]{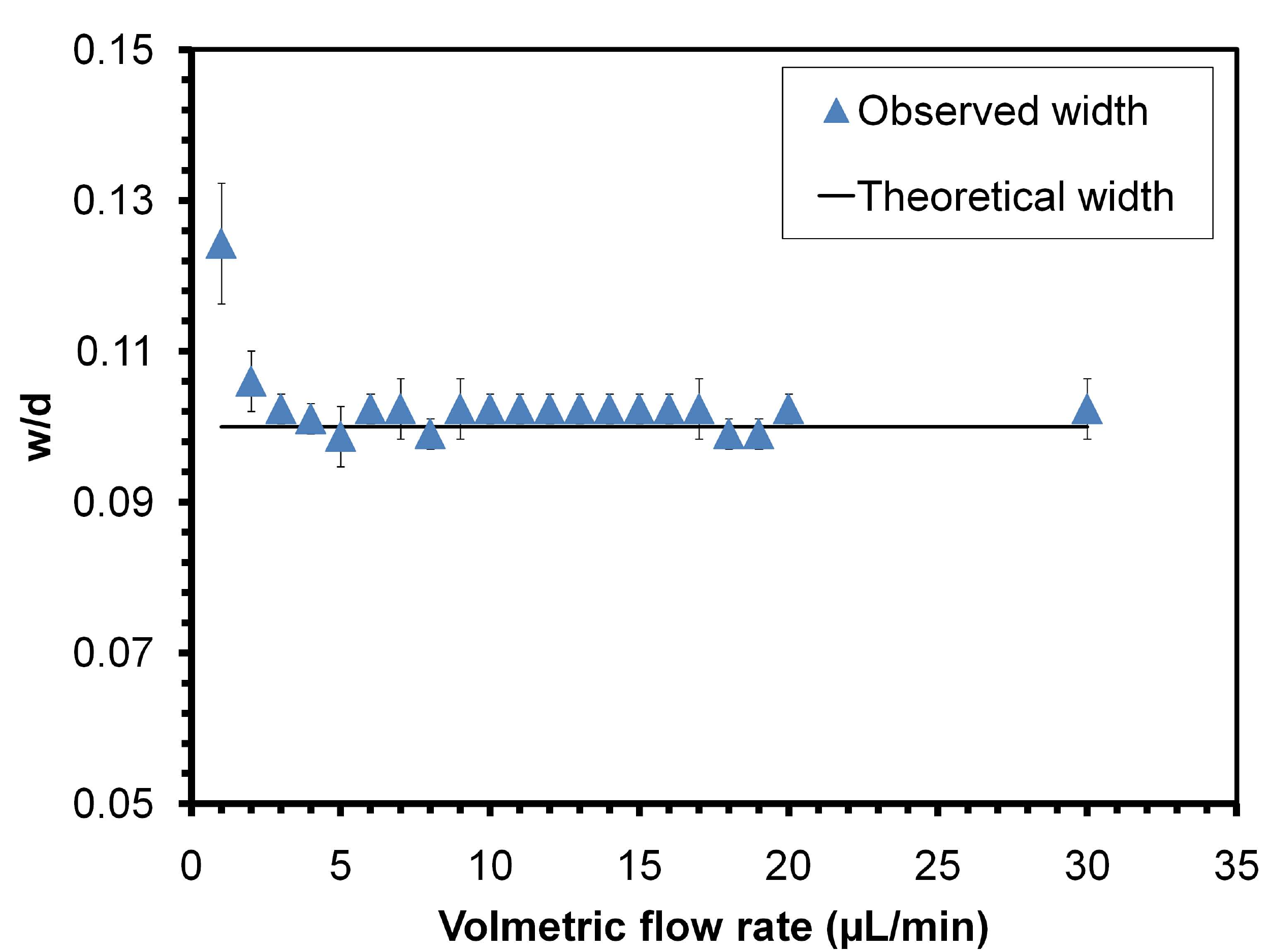}
\caption{Validation of the width determination. Here the flow rate
  ratio is kept constant at $\phi$=50, increasing the total flow
  rate.}
\label{fig:ConstChangingFRRplot}
\end{figure}

To study the nature of flow, we carry out the experiments in the
absence of lipids, thereby not including the additional complexity of
liposome formation. Instead, we introduce a dye---filtered fountain
pen ink---to enable visualisation.  Experiments were conducted at
various flow rate ratios identical to that employed in
\Figref{fig:phivar}. Since the dye is introduced in the core fluid, we
expect that in absence of diffusion of dye, or any other convective
instability, the width $w$ of the core (coloured) region in a fully
developed flow would be
\begin{equation}
  \label{eq:wphi}
   w \approx \frac{d}{\sqrt{2 \, \phi}}
\end{equation}
This can be obtained by integrating the fully developed Poiseuille
flow parabolic velocity profile (assuming uniform viscosity) for the
core and annular-fluid volumetric flow rates, and in the limit of $w
\ll d$.  We validate the experimental method of determining the width
of the dye-region (core fluid) which results in the data shown in
\Figref{fig:ConstChangingFRRplot}. We choose a large value of $\phi =
50$, and vary the total flow rate. Here, $17 < \Pe < 507$ and $ 2 <
\Rey < 64$. As expected, the diffusive mixing should be minimal for
large \Pe, and the observed width is in agreement with the expected
value from \Eqref{eq:wphi}.

\begin{figure}[tb]
\begin{center}
\includegraphics[width=0.9\linewidth]{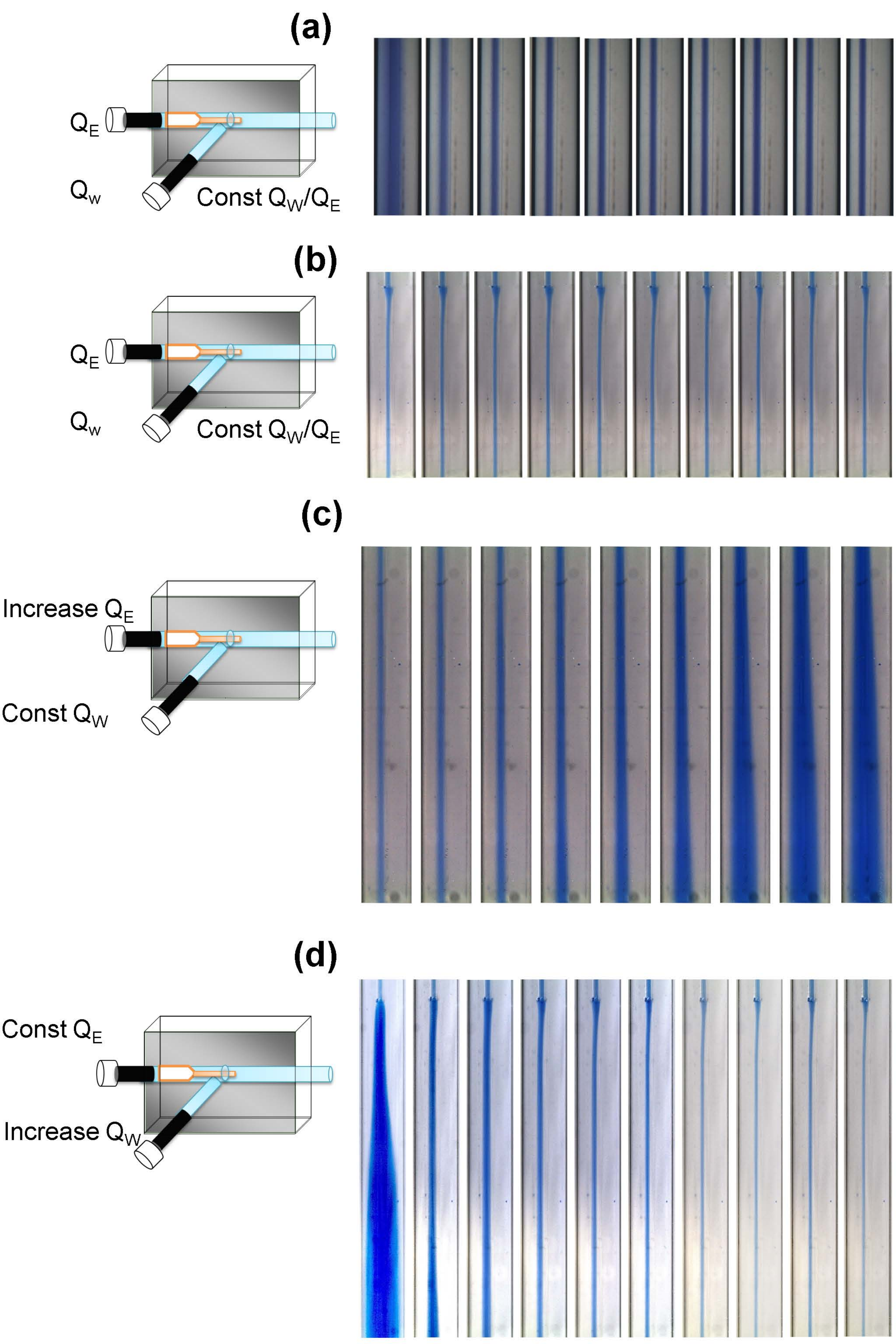}
\caption{Visualisation of mixing of a blue dyed ethanol (without the
  lipids) with water, in four typical cases of flow rates employed for
  liposome synthesis in \Figsref{fig:phivar} and~\ref{fig:Qtot}.  (a)
  Low $\phi=20$ kept constant, and total flow rate $Q$ is varied such
  that $Q_{\mathrm{E}}=5,10,\ldots,45,50$~$\mu$l/min, going from left
  to right. The photographs shows wider region of mixing at lower
  $Q$. (b) High $\phi=50$ kept constant, and total flow rate $Q$ is
  varied such that $Q_{\mathrm{E}}=2,4,\ldots,16,\&30$~$\mu$l/min,
  going from left to right. The photographs show no mixing of the
  dye. (c) Constant $Q_{\mathrm{W}} = 100~\mu$l/min and increasing
  $Q_{\mathrm{E}}=1,2,\ldots,8,9~\mu$l/min showing strong mixing for
  low $\phi$ (right most). (d) Constant $Q_{\mathrm{E}}=5~\mu$l/min
  and increasing $Q_{\mathrm{W}} =
  100,200,\ldots,1000~\mu$l/min. showing strong mixing at low $\phi$
  (left most).}
\label{fig:differentHDF}
\end{center}
\end{figure}

We have considered a similar set of flow rates as used in the liposome
synthesis in \Figref{fig:phivar}. In \Figref{fig:differentHDF} we have
considered three variations: (a) Constant flow rate ratio $\phi = 20$
and $\phi = 50$, (b) Varying $\phi$ keeping outer flow rate constant,
and (c) Varying $\phi$ keeping the inner flow rate constant.  At a
large flow rate ratio of $\phi = 50$, we see that the width of the
core region remains the same.

\begin{figure}[tb]
  \centering
  \includegraphics[width=0.9\linewidth]{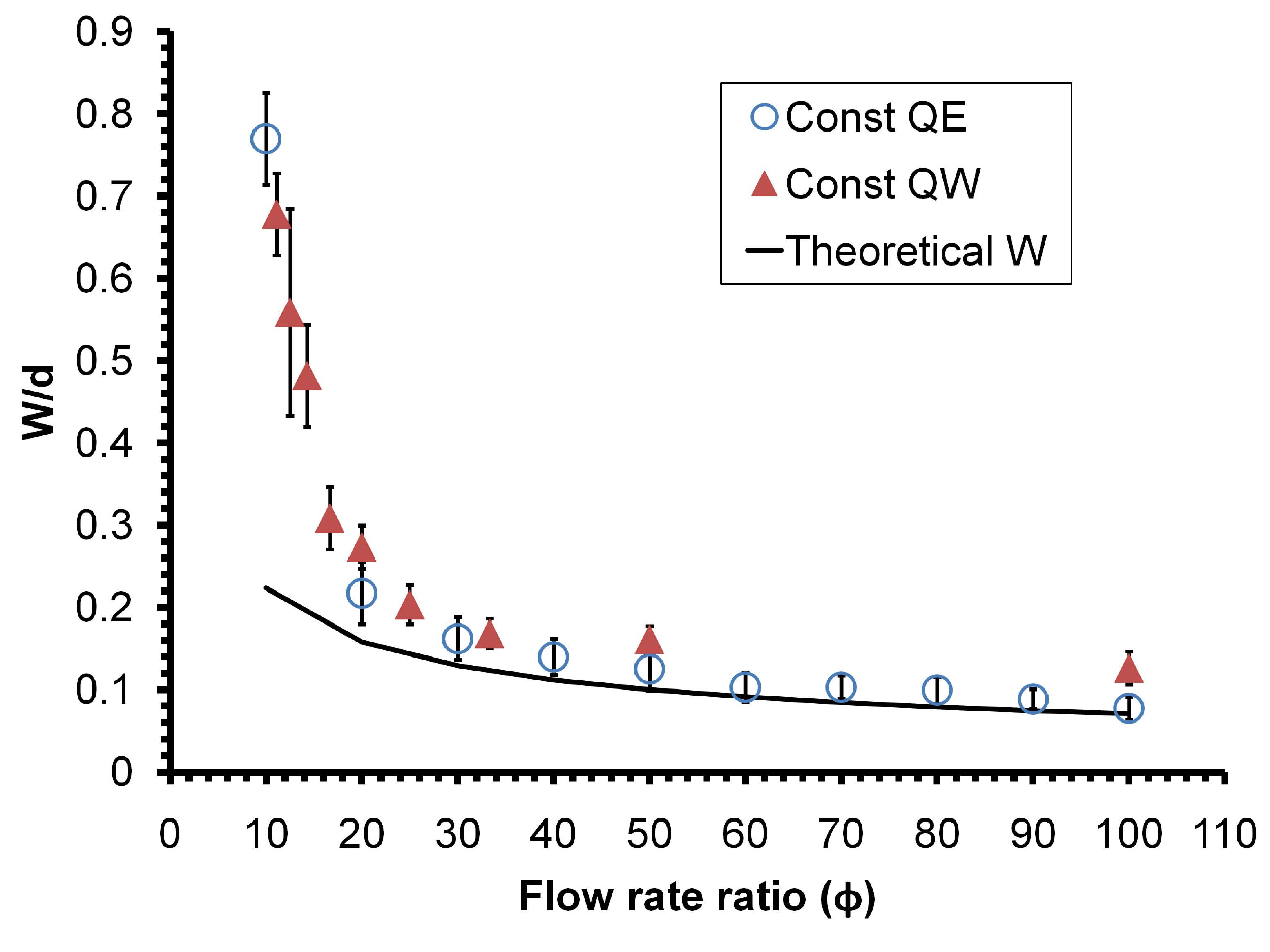}
  \caption{Comparison of the observed spread of the dyed ethanol with
    near the outlet of the channel, along with the theoretical
    approximation of the width obtained from
    \Eqref{eq:wphi}.}
  \label{fig:wcomp}
\end{figure}

In \Figref{fig:differentHDF}~(c) we see an unexpected behaviour.  The
width of the dye region increases with increase in the inner flow
rate. We expect that this will decrease $\phi$ from about 100 to about
10.  Though this implies an increase in $w$ from \Eqref{eq:wphi}, the
increase is much more than what is theoretically expected, as shown in
\Figref{fig:wcomp}.  In this set of experiments the Peclet number is
high at $\Pe \approx 35$, and increases only marginally from \Pe=33 to
\Pe=36, so this increase in the width cannot be explained even by
Taylor diffusion of ethanol into water.  Moreover, the dye particles
being larger than ethanol, the Peclet number for dye particles is even
larger at $\Pe_{\mathrm{dye}} > 10^3$.  Therefore, the explanation of
the increase in the observed width cannot be given based on mere mass
conservation or diffusion, and we conclude that a gentle convective
mixing of the fluids takes place due to the growth of an unstable
mode. This is a gentle mixing because by nature of the flow, once the
fluids mix at low \Rey\ the viscosity becomes uniform, which
stabilises the flow.

A similar situation is seen when the outer flow rate is varied keeping
the inner flow rate constant, as seen in \Figref{fig:differentHDF}~(d)
Here too the Peclet number is high, $36 < \Pe < 335$.  While we can
conclude the existence of a convective instability that leads to
mixing of the fluids, it is difficult to correlate this to the
theoretically expected instability value of the \Pe\ or the \Rey, as
the specific model considered in \cite{Selvametal07} is different from
the present system.

\begin{figure}[tb]
\begin{center}
\includegraphics[width=0.9\linewidth]{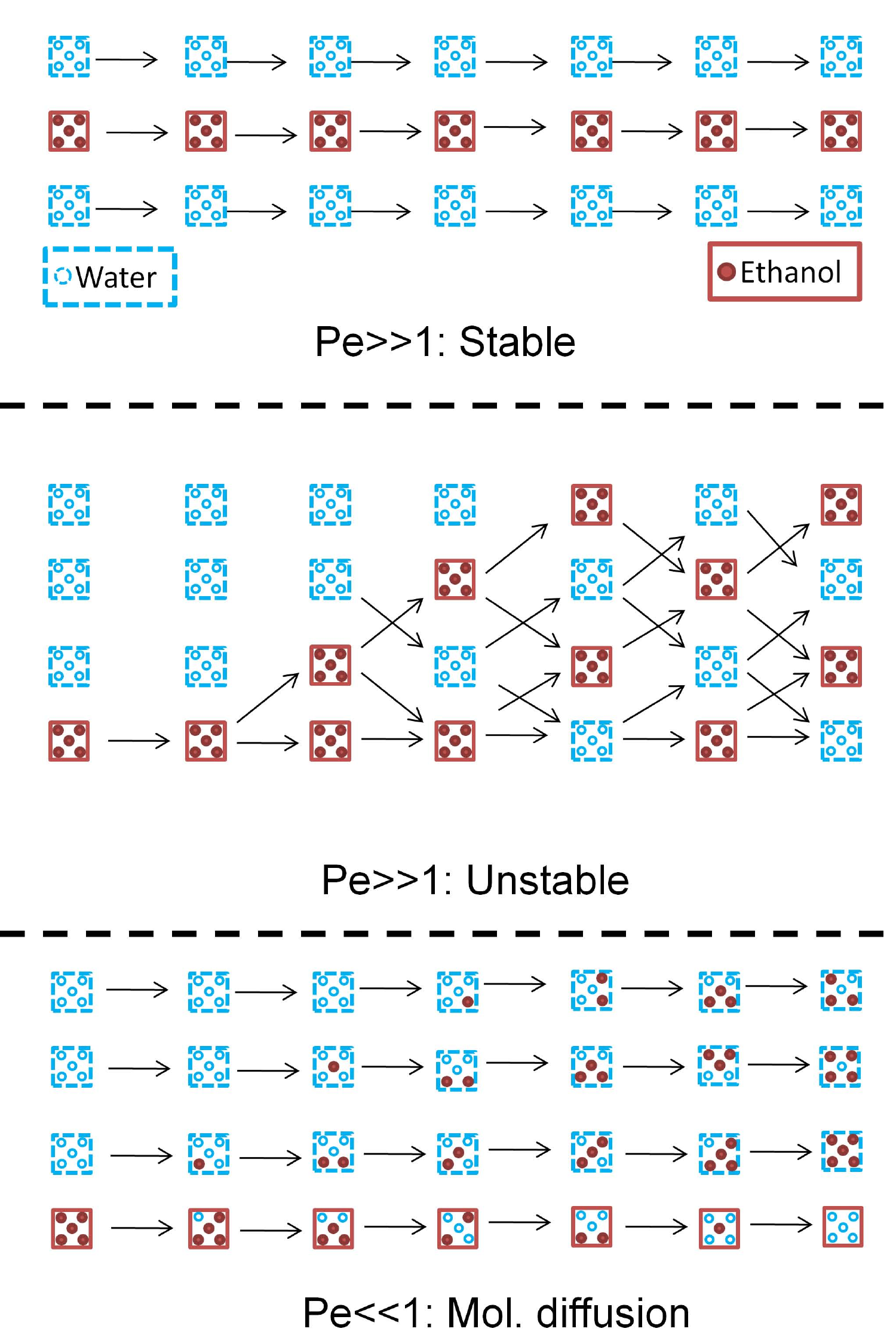}
\caption{Schematic representation of various forms of mixing observed
  in \Figref{fig:differentHDF}. Here a box denotes a packet of fluid,
  and light (blue) circles denote water molecules, and dark (red)
  circles denote ethanol (dyed) molecules. Top: High Peclet, $\Pe \gg
  1$, without any instability the packets traverse in a laminar flow
  without mixing, as in the case of $\phi=50$ in
  \Figref{fig:differentHDF}. Middle: Hydrodynamic instability driven
  mixing of fluids in high $\Pe \gg 1$ implies the particles
  are still inside the packet of fluid, but the packets themselves mix
  spreading out ethanol (dye). Bottom: In low $\Pe \ll 1$ dominated by
  diffusion, the spread of ethanol (dye) is due to exchange of
  particles in the packets, while the packets themselves traverse in a
  laminar flow.}
\label{fig:MF_mixing_cartoon}
\end{center}
\end{figure}

The mixing of the fluids can be schematically explained with the help
of \Figref{fig:MF_mixing_cartoon}. In the stable case with high \Pe, the
fluid packets move in a laminar fashion, and the diffusion of the
particles out of the packet is negligible.  Whereas, when the
viscosity stratification leads to an instability, the flow does not
remain laminar---it mixes the packets of fluids. Here too since the
\Pe\ is large, the diffusion of particles out of the packet is
negligible on the length scale of the channel diameter, however when
seen from a macroscopic perspective, the dye particles occupy a wider
region in space as the packets have themselves been dispersed by
mixing.  To contrast this, a diffusive mixing occurs at a low Peclet
and laminar flow, where packets of fluid exchange particles from other
packets, purely by diffusion.

\subsection{Droplet mixing regime}
\begin{figure}[tb]
\begin{center}
  \centering
\includegraphics[width=0.9\linewidth]{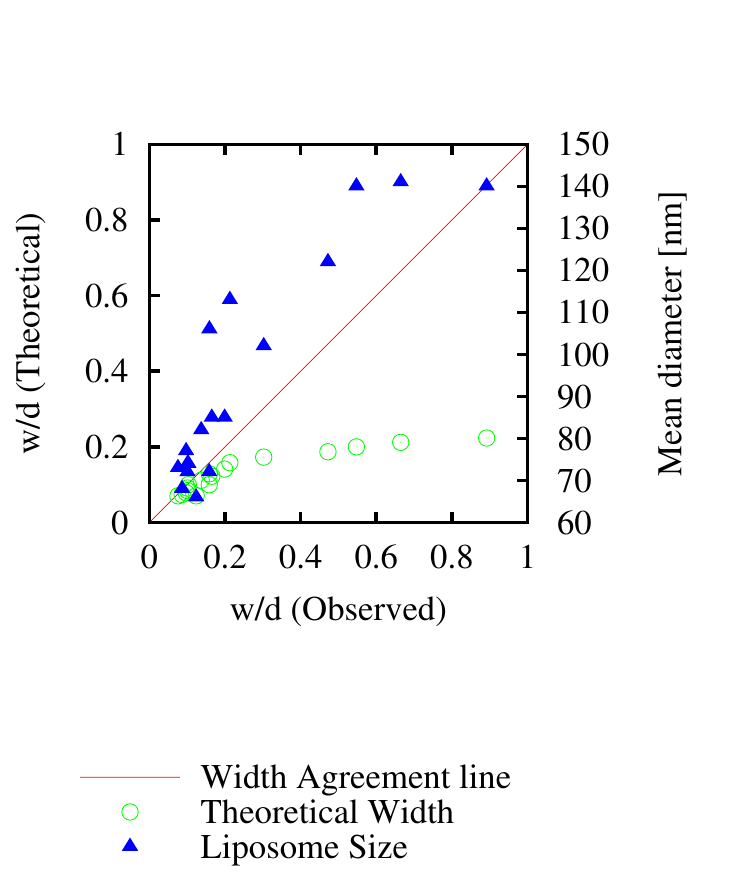}
\caption{Correlation of the mixing observed visually in
  \Figref{fig:wcomp} with the liposome size obtained under identical
  flow conditions. Liposome size are smaller when the theoretical
  width tends to the observed (droplet mixing). Liposomes of
  largest size are formed when there is a large disagreement in the
  widths (convective mixing). }
\label{fig:comparison-w}
\end{center}
\end{figure}

From the above observations we conclude that not all high \Pe\ and low
\Rey\ cases leads to an instability. There are regions where there is
an agreement of the theoretical width with what is observed.  This is
summarised in \Figref{fig:comparison-w}, where we see that for very
small $w$ there is close agreement, whereas it begins to deviate at
$w/d$ around 0.3. In the regime of agreement, the width of the core
flow remains nearly constant throughout the length of the channel. At
the outlet of the channel, the flow forms droplets.  Since the two
streams remain unmixed inside the channel, even while forming the
drops we observe two visually distinct regions.  The fluids only mix
when the droplets are gathered in the outlet reservoir (cuvette).

In \Figref{fig:comparison-w}, the correlation between the liposome
size and the width of the dye can be observed.  Though these are two
different experiments, they have been carried out under the same flow
rate conditions.  The liposome size is plotted against the observed
width, each obtained under the same flow rate values. This plot can be
used to infer the nature of flow that would have occurred in the
liposome synthesis experiment.  When the theoretical width tends to
agree with the observed width, the liposome size is on the lower side,
implying the liposome size is governed more by the droplet mixing
regime. Moreover, when the observed width deviates from the theory,
the liposome size is the largest.  This implies that it is the
internal convective mixing that leads to a larger size.

\subsection{A plausible explanation for liposome size selection}

To summarise the findings, liposomes are smaller when the streams mix
as droplets when collected in the cuvette, and are larger and when the
streams mix by convection inside the channel.  The size of liposomes
formed in the droplet mixing regime are also comparable with those
obtained in the ethanol injection and stream mixing regimes shown in
\Figref{fig:different-mixing}.

This can be explained if we assume that the size of the liposomes
depends on the nature, and hence the length scale of convective
mixing. As water mixes with ethanol, the lipids self-assemble when it
is no longer favorable to remain in a dispersed form. They begin to
assemble as disc-like bilayer membrane micelles \citep{Leng20031624}
which grow in size.  When the edge energy at the periphery of the
discs become comparable to thermal fluctuations induced bending energy
of the membrane, the discs close upon to form vesicles
\citep{Helfrich1973693}.  The presence of ethanol essentially
reduces the interfacial tension along the edges.  A reduced tension
will lead to a larger sized liposome, since the thermally induced
bending of the membrane would take place at a larger size of the discs.
Left to itself all the vesicles formed would be about the same size as
determined by this balance.  However, convective motion in the length
scales of the discs can induce closure before the membrane grows to
larger possible dimensions.

Turbulent mixing (droplet mixing, ethanol injection, and stream
mixing) has smaller length scales of motion, thereby leading to
smaller vesicles. It is known that the smallest vesicles are induced
by the action of ultrasonic waves, that essentially introduce
convective motion in small length scales by spontaneous vapourisation
into bubbles \citep{Richardson20074100, Yamaguchi200958}.  The
instability introduced by viscosity stratification leads to a rather
gentle mixing, because mixing makes the viscosity uniform, and this
suppresses the instability (a negative feedback). In these cases, the
length scales are possibly higher than in turbulent mixing, and hence
the size of liposomes are larger.

\section{Conclusion}

The microfluidic method employed here for liposome synthesis is a
variant of the ethanol injection technique in which lipids dissolved
in ethanol is contacted with water, leading to the assembly of lipids
to closed vesicular structures in the presence of excess water.  Two
distinct but connected aspects of microfluidic synthesis of liposomes
were shown.

The first aspect is a demonstration of a simple device capable to
synthesise unilamellar liposomes of nearly uniform size in a co-axial
core-annular flow geometry.  Detailed observations of the effect of
various flow parameters, lipid chemistry and concentration were made.
The chief conclusion from these observations are that the size and
formation of liposomes cannot be characterised by ternary component
diagrams, or just the flow rate ratio $\phi$ as done in the past; and
that the size strongly depends on the nature of flow.

The second and more significant aspect is on throwing some light on
the hitherto unknown mechanism of liposome size selection in ethanol
injection techniques and by implication in a microfluidic setup such
as the one employed here.  We showed that the size depends strongly on
the nature of micro-convective mixing.

In the case of the microfluidic set up, there are two regimes. One in
which a convective hydrodynamic instability due to viscosity
stratification leads to a gentle mixing and a larger size of
liposomes.  In the other regime the streams do not mix inside the
channel, but mix as droplets formed at the outlet when collected in a
reservoir. Vigorous mixing leads to smaller liposomes.  Though we do
not present a detailed theory for the exact value of size selection,
it is argued that it is the length scales of flow (mixing) that
determine the size of liposomes.
 
Further studies into the detailed structure of the flow in
microfluidic and other ethanol-injection methods can establish a more
accurate connection between the liposome size and the length scales of
flow variation.  The discovery of this important effect should help in
formulating models that will be able to predict liposome size and
lamellarity in many coacervation methods.  Apart from flow, other
aspects of lipid concentration, membrane composition, effect of
alcohol on the bending modulus \citep{LyLongo2004}, temperature etc.,
also require attention, but needs to be carried out in conditions where the
influence of flow can be made negligible.

\paragraph{Acknowledgement}

The financial support of Department of Science and Technology,
Government of India, through a special IRHPA grant, is gratefully
acknowledged. Imaging Facilities in SAIF and CRNTS, IIT Bombay were
employed for liposome visualisation.  We thank FEI company for
providing us free analysis of our samples in their Quanta 250 FEG
ESEM. PS also wishes to thank DuPont, Wilmington, for the Young
Professor fellowship that supported a part of this work.

\section*{References}
\bibliographystyle{model2-names}
\bibliography{liposomes}
\end{document}